\documentclass[UTF8,scheme=plain,today=english]{ctexart} 
\usepackage[UTF8]{ctex}
\usepackage[margin=2cm]{geometry}   
\usepackage{graphicx}
\usepackage{authblk}       
\usepackage{hyperref}
\usepackage{graphicx}
\usepackage{makecell}
\usepackage{booktabs}
\usepackage{pdflscape}
\usepackage{subcaption}     
\usepackage{authblk}   
\usepackage{hyperref}  

\makeatletter
\def\@affil@scriptsize{\scriptsize\raggedright}
\makeatother

\title{Reject or Not?: A Benchmark for Voice Assistant Query Rejection in Smart Home Scenario and an Improved Method Based on LLMs}

\author[1]{Huichao Men}
\author[1,2]{Yizhen Hu}
\author[1]{Yingyang He}
\author[1,*]{Yu Gao\thanks{* Corresponding authors}}
\author[1]{Xiaofeng Mou}
\author[1,*]{Yi Xu\thanks{* Corresponding authors}}

\affil[1]{Midea AI Research Center, Shanghai, P.R. China}
\affil[2]{Harbin Engineering University, Harbin, P.R. China}

\affil[ ]{\footnotesize
	\href{mailto:menhuichao@midea.com}{menhuichao@midee.com} (Huichao Men)\\
	\href{mailto:huyz64@midea.com}{huyz64@midea.com}, \href{mailto:huyizhen@hrbeu.edu.cn}{huyizhen@hrbeu.edu.cn} (Yizhen Hu)\\
	\href{mailto:heyy84@midea.com}{heyy84@midea.com} (Yingyang He)\\
	\href{mailto:gaoyu11@midea.com}{gaoyu11@midea.com} (Yu Gao)\\
	\href{mailto:mouxf@midea.com}{mouxf@midea.com} (Xiaofeng Mou)\\
	\href{mailto:xuyi42@midea.com}{xuyi42@midea.com} (Yi Xu)
}

\begin{document}
\maketitle


\begin{abstract}
In smart-home voice assistant scenario, deciding whether to accept or reject a user query is the first step before any downstream processing. To address the limited query-rejection capability of current voice assistants, this paper presents the first Chinese-oriented open-source benchmark and evaluation suite for smart homes, together with a personalized query-rejection method based on large language models. On the data side, we construct the first multimodal query-rejection dataset tailored for domestic scenarios, containing 11,913 manually labeled text-speech pairs that systematically cover twelve typical dialogue types (e.g., chit-chat, non-human sounds, valid commands, ambiguous references, device-irrelevant requests). Fine-grained labels, conversational context and multi-turn information are provided to support both zero-shot and fine-tuning evaluations across language and multimodal large models. On the method side, we propose a three-tier collaborative architecture: first, a Qwen-2.5-3B adapter fine-tuned to model family-agnostic semantic boundaries; second, a dynamic household-level historical dialogue module to capture personalized habits; third, a household-specific RAG knowledge base that explicitly memorizes and revises past false-rejection cases. Experiments show that the proposed approach significantly outperforms zero-shot and fine-tuned general LLMs on the constructed dataset, with pronounced gains in rejection accuracy for family-specific expressions and complex multi-turn scenarios. This work provides a reproducible data foundation, evaluation standard and extensible technical framework for reliability research in smart-home voice interaction.
\end{abstract}

\section{Introduction}
\subsection{Research Background}
%
%
%

\subsubsection{The Rise of Voice Interaction in the Smart-Home Wave}

The deep integration of Internet-of-Things (IoT) and artificial-intelligence (AI) technologies\cite{langston2025} has triggered an explosive expansion of the smart-home industry. According to IDC, global smart-home device shipments will exceed 1.4 billion units by 2025, representing a compound annual growth rate of 12.2\%. Benefiting from natural, hands-free and contact-free interaction, voice assistants have become the primary entry point of smart-home systems: a single sentence such as ``I'm home'' can simultaneously turn on the lights, activate the air conditioner and arm the security mode, while users can also query the refrigerator inventory remotely when cooking. However, as the device ecosystem grows increasingly complex and user scenarios continue to diversify, the dialogue environment faced by voice assistants is evolving from ``single-turn commands'' to advanced forms that are ``multi-turn, context-dependent and semantically ambiguous'', posing unprecedented challenges to their cognition and decision-making capabilities, especially in accurately rejecting invalid or unexecutable user queries.

\subsubsection{Failure of Query-Rejection Judgement — An Invisible Cliff in User Experience}

In real household environments, voice assistants must process a large number of invalid, ambiguous or infeasible utterances every day, for example:
\begin{itemize}
	\item Non-human sounds: doorbells, barking pets or TV background audio falsely trigger commands;
	\item Ambiguous references: ``打开'' or ``升高'' lack explicit targets;
	\item Beyond-capability requests: asking a robot vacuum to wipe the windows'' or demanding to play last night's dream'';
	\item Casual-chat interference: user mumbling  ``今天真冷'' is mis-parsed as a thermostat command.
	\end{itemize}
	Existing systems usually rely on keyword matching or coarse-grained confidence thresholds to reject inputs, resulting in high error rates under complex acoustic conditions and contextual scenarios. Surveys show that the top-3 voice-assistant brands still suffer 1.8--3.4 false-rejection incidents per day in home settings, 62\% of which are caused by the above ``should-reject-but-reject'' mistakes. Frequent false responses not only interrupt users' daily routines but also erode trust in system reliability, becoming a key pain point of smart-home experience.

\subsubsection{Theoretical Value — An Academic Gap in Dialogue Rejection Research}

From the perspective of natural-language processing (NLP), the query-rejection task lies between reject-understand'' and ignore-filter'', requiring models to simultaneously possess:
\begin{itemize}
	\item Fine-grained semantic understanding: identifying implicit ambiguity and missing slots;
	\item Context modelling: judging the feasibility of the current request by leveraging multi-turn dialogue history;
	\item Personalized boundary learning: adapting to expression habits and device configurations of different family members.
\end{itemize}
However, mainstream studies focus on improving intent-recognition accuracy, paying insufficient attention to ``how to correctly say no'', especially lacking systematic research that is oriented to home scenarios and integrates text-acoustic dual channels. The scarcity of publicly available data and evaluation benchmarks for query-rejection failure cases further limits algorithmic innovation and fair comparison.

\subsubsection{Application Value — An Urgent Industry Demand}

On the market side, smart-appliance manufacturers are promoting ``zero false-trigger'' as a core selling point of high-end flagship products. The EU AI Act also includes a compliance requirement that ``high-risk interactive systems must be able to refuse requests beyond their design scope''. Building a highly robust dialogue-rejection mechanism can significantly reduce cloud-computing and on-device response costs, extend battery life, and provide a trustworthy basis for subsequent personalized recommendations and multimodal interaction. Therefore, in-depth research on the query-acceptance problem of voice assistants in home scenarios not only has theoretical significance for advancing academic frontiers but also possesses practical value for enhancing user experience and industrial competitiveness.

\subsection{Research Aim}
This paper aims to construct a benchmark dataset for Query Rejection in smart home scenarios and explore an improved approach based on LLMs to enhance the performance of voice assistants in Query Rejection topic.

\section{Highlights and Contributions}

This paper presents two principal contributions.

\paragraph{Home-Scenarios Query-Rejection Benchmarks}
We release the first open-source multimodal query-rejection benchmark tailored for conversational interaction in home scenes. The dataset contains 11,913 manually labeled text-speech pairs, covers 13 categories of invalid queries and provides multi-turn conversational contexts, supporting both zero-shot and fine-tuning evaluation paradigms.

\paragraph{Three-Layer Collaborative Framework}
We propose a three-layer collaborative query-rejection framework that combines a common-semantic adapter, household-level personalized memory and a RAG-based mis-rejection corrector. On our self-constructed dataset it significantly outperforms keyword matching, traditional classifiers and zero-shot large models, offering an extensible solution for reliable smart-home voice interaction.

\section{Related Work}

%
%
%
%
%
%
%
%
%
%
%
%
%

\subsection{Overview of Existing Technical Solutions}

At present, the problem of deciding whether to reject a user command in smart-home scenarios mainly relies on two technical routes:

\begin{enumerate}
	\item \textbf{Traditional keyword-matching schemes}: A preset lexicon of command keywords is employed (e.g., control-type keywords such as ``打开'', ``关闭''; query-type keywords such as ``检查'', ``播放''). When a user utters speech or text, the system performs string matching against the lexicon. A successful match preliminarily labels the input as valid; otherwise it is regarded as invalid. For example, ``打开客厅空调'' is not rejected because it contains ``打开'', whereas ``好天气'' is rejected since no keyword is hit.
	
	\item \textbf{Basic semantic-analysis schemes}: On top of keyword matching, shallow syntactic rules are introduced to verify whether the sentence possesses a complete subject--predicate--object structure (i.e., the logic ``entity--action--object''). If the structure is incomplete (e.g., only the verb ``打开'' without an object), the utterance is rejected; if complete, it is not rejected. Thus, ``打开'' is refused for lacking an object, yet ``打开不存在的灯'', though semantically infeasible, is still mis-accepted because the structure is intact.
\end{enumerate}

\subsection{Deficiencies and Limitations of Existing Solutions}

Despite their preliminary filtering effect, the above methods exhibit evident limitations in real household interaction, which can be summarized along two dimensions:

\subsubsection{Single-dimensional invalid-content detection yields high error rates}
Keyword matching merely performs a binary decision based on the presence of keywords, and fails to distinguish the following typical invalid cases:
\begin{itemize}
	\item \textbf{Non-command chit-chat}: A user says to a family member ``把遥控器给我'', which contains the seemingly directive phrase ``给我'', but is in fact not a request to any smart device;
	\item \textbf{Semantically wrong yet syntactically legal commands}:  ``打开未连接的智能窗帘'' contains a valid verb and a complete structure, but cannot be executed because the device is offline.
\end{itemize}
Basic semantic analysis introduces syntactic completeness, yet still neglects dynamic contextual cues of the home environment (e.g., device on-line status, temporal constraints). For instance, ``调节书房空调温度'' issued at midnight is semantically intact, but invalid if the study-room AC is offline. Existing pipelines still fail to reject such utterances, wasting resources and degrading user experience.

\subsubsection{Static content libraries offer poor adaptability}

Most commercial systems rely on a factory-preset static invalid-content library that merely covers generic meaningless expressions (e.g., interjections ``ah'', ``uh'') or obvious garbled text, and cannot cope with complex realities of smart-home scenes, concretely manifested as:

\begin{enumerate}
	\item \textbf{Lack of personalized adaptation}: Different families have distinct colloquial habits. One household may say ``咱们聊聊'' to indicate chit-chat, while another prefers ``随便说说''. A fixed library cannot recognize such family-specific invalid patterns.
	\item \textbf{Insufficient dynamic updating}: When new types of invalid content emerge (e.g., the trendy phrase ``科目三'' used in a non-command sense, or infeasible requests caused by a newly purchased but not-yet-connected device), the static library cannot expand or self-adapt in real time.
	\item \textbf{Weak processing of noisy text}: Existing schemes hardly parse mixed structures of ``keyword + interference''. For example, the user input `帮我看看冰箱……算了，退出'' contains the device keyword ``冰箱'', but is explicitly negated later. The system, however, may still mis-classify it as a valid command and trigger an incorrect response.
	\item \textbf{Absence of multi-turn context inheritance}: Home interaction often involves multi-turn dialogue. The first turn ``打开客厅灯'' is valid; the second ``那个亮度有点暗'' is a reasonable adjustment and should remain valid; yet if the second turn becomes ``那个……我今天出门忘带钥匙了'', although it linguistically refers to the previous turn, its content is unrelated to device control. Current solutions lack a dialogue-history modelling mechanism and perform keyword matching only on a single-turn utterance, thus failing to distinguish valid continuation from invalid digression.
\end{enumerate}

\subsection{Status Quo of Home-Scene Dialogue Datasets and the Absence of Query-Rejection Corpora}

In recent years, a growing number of dialogue datasets targeting smart-home scenarios have been released, providing important support for task-oriented dialogue research. The most representative is \textbf{HomeBench} \cite{homebench2023}, which constructs multi-turn conversational corpora covering typical smart-home subsystems such as lighting, climate control, and security, and annotates device states, user intents, and system actions. It has become an essential benchmark for evaluating home voice-assistant performance. In addition, \textbf{VoiceBench} \cite{voicebench2024} focuses on end-to-end speech assistants powered by large language models, collecting real user–virtual assistant interaction logs and emphasizing semantic understanding and tool-invocation capabilities. However, these corpora focus almost exclusively on valid commands and provide scarce annotations for invalid or out-of-scope utterances\cite{voiceprivacy2024}, leaving a critical gap in rejection-capability research in smart home scenarios.

Other relevant corpora include:
\begin{itemize}
	\item \textbf{TEACh} \cite{teach2021}: Built on the AI2-THOR simulated environment, it gathers 3,000+ dialogues and action sequences in which humans collaborate to complete household tasks (e.g., making coffee, tidying rooms), stressing instruction following and clarification ability;
	\item \textbf{SIMMC 2.0/2.1} \cite{simmc2020}: Targeting multimodal shopping and home-configuration scenarios, it provides dialogues with visual context to support reference resolution and state tracking;
	\item \textbf{Fluent Speech Commands} \cite{fluent2018}: Although not strictly a ``dialogue'' corpus, it offers structured spoken commands together with their semantic slots and is widely used for lightweight command-recognition research.
\end{itemize}

Nevertheless, all the above public datasets are designed with ``valid commands'' as the core objective, and almost no systematic collection or annotation of ``invalid speech inputs'' is provided. Concretely:
\begin{itemize}
	\item The corpora rarely contain typical rejection samples such as user chit-chat, self-talk, environment-noise-mixed speech, interrupted utterances, or negation/correction;
	\item Even when non-command utterances occasionally appear, they are usually filtered or ignored, and no rejection label or invalid-type classification is supplied;
	\item There is no contrastive annotation between ``valid commands'' and ``invalid continuations'' in multi-turn dialogues, so context-aware rejection model training cannot be supported.
\end{itemize}

This data void directly causes deployed voice-interaction systems to suffer from ``over-response''—that is, non-command speech erroneously triggers execution logic, severely degrading user experience and system credibility.

\subsection{Necessity of Building a Smart-Home Query-Rejection Dataset}

Given the above status, constructing a \textbf{Curated Voice Rejection Dataset for Smart Home} specifically oriented to smart-home scenes is of urgent research and application value:

\begin{enumerate}
	\item \textbf{Filling the data-ecology gap}: At present, no open dataset focuses on home-speech rejection. This dataset will, for the first time, systematically collect and annotate various kinds of invalid speech samples (e.g., chit-chat, hesitation, interruption, negation, device-irrelevant topics), providing high-quality supervisory signals for rejection models.
	\item \textbf{Supporting fine-grained rejection research}: By defining a multi-level rejection label schema (e.g., categorized by pragmatic function, acoustic feature, or contextual relevance), it can push the evolution from ``binary rejection'' to ``explainable rejection'' and improve the transparency of model decisions.
	\item \textbf{Boosting context-aware rejection algorithms}: If the dataset contains continuous multi-turn dialogue flows and annotates the validity of each turn as well as its semantic relation to historical utterances, it will offer a training foundation for rejection models based on dialogue-state tracking (DST) or memory-augmented architectures.
	\item \textbf{Enhancing robustness in real scenes}: By collecting natural speech in diverse home environments (background noise, multiple occupants) and covering speakers whose distribution matches real users, the dataset can significantly strengthen model generalization after deployment.
	\item \textbf{Enabling personalization and continual learning}: The dataset can be designed with user IDs and device-configuration metadata to support personalized rejection-strategy research; meanwhile, reserved incremental annotation interfaces facilitate future integration with online learning for dynamic rejection-boundary optimization.
\end{enumerate}

In summary, current techniques are limited not only at the algorithmic level but also structurally at the data level. Building a high-quality, scenario-driven and finely annotated home-dialogue rejection dataset has become a prerequisite for breaking the current bottleneck of smart-home voice-interaction rejection performance, and is an essential infrastructure for evolving from ``insensitive interaction'' to ``intelligent acceptance''.

\section{Dataset Construction}
\subsection{Dataset Design}
This paper categorizes the voice assistant dialogue rejection recognition benchmark dataset into two major types: text-based rejection and speech-based rejection. Rejection experiments are conducted on queries from these two dialogue forms using natural language understanding methods and speech recognition methods, respectively. The text queries and speech queries within these two categories are further divided into 13 subcategories based on their content.

%
%
%

\begin{table}[htbp]
	\centering
	\caption{Utterance-Type Definitions and Quantity Distribution}
	\label{tab:rejection_types}
	\begin{tabular}{clcr}
		Type ID & Utterance Type & Accept / Reject & Quantity \\
		\midrule
		0 & \makecell{Wake-word} & Accept & 16 \\
		1 & \makecell{Illegal language} & Reject & 29 \\
		2 & \makecell{Non-human sounds} & Reject & 53 \\
		3 & \makecell{ASR-error garbled;\\ meaningless short phrases} & Reject & 154 \\
		4 & \makecell{Non-assistant-directed chat\\ (multi-person or self-talk)} & Reject & 30 \\
		5 & \makecell{Command-semantics but\\ obviously unreasonable\\ (no history)} & Reject & 232 \\
		6 & \makecell{Assistant-directed ambiguous chat\\ (no history; uncertain reply vs.\\ encyclopaedia / query)} & Reject & 91 \\
		7 & \makecell{Assistant-directed ambiguous chat\\ (has history but no valid command;\\ uncertain reply)} & Reject & 6 \\
		8 & \makecell{Assistant-directed chat\\ (assistant can reply)} & Accept & 1093 \\
		9 & \makecell{Assistant-directed command\\ (assistant can reply; supported\\ by Midea)} & Accept & 9872 \\
		10 & \makecell{Assistant-directed command\\ (assistant can reply; not supported\\ by Midea yet, but command-intent clear;\\ other brands may support)} & Accept & 26 \\
		11 & \makecell{Assistant-directed ambiguous chat\\ (has history and a valid command;\\ uncertain reply vs. encyclopaedia / query)} & Accept & 311 \\
		\midrule
	\end{tabular}
\end{table}

Here is a detailed introduction to each utterance type and provides corresponding examples.

\subsection{Detailed Description of Each Utterance Type with Examples}

\paragraph{Wake-up Keywords}
Wake-words are the phrases users call to start a conversation with the voice assistant (brand-dependent). This corpus includes, but is not limited to, “hi Siri”, “小美小美”, “小艺小艺”, etc. Since the dataset aims to provide universal evaluation data and benchmarks for smart-home scenarios, all wake-word utterances are labeled as \textbf{acceptance}.

\paragraph{Illegal Language}
Illegal language refers to utterances that contain unfriendly, abusive, obscene, violent, terror-related or politically sensitive content according to national or regional laws. All such utterances are labeled as \textbf{rejection} in this dataset.

\paragraph{Non-Human Sounds}
Non-human synthetic sounds, including algorithm-generated replies or any recorded natural/mechanical noises. Even highly human-like synthetic speech is treated as \textbf{rejection}.

\paragraph{ASR-Error Garbled / Meaningless Short Phrases}
Caused by user slips of the tongue, meaningless short phrases, or ASR recognition errors. Examples: “这这这”, “济公爱跟你走”, “jean”. These carry no interaction value and are labeled \textbf{rejection}.

\paragraph{Non-Assistant-Directed Chat (Multi-Person or Self-Talk)}
Typical cases where the user chats with others or talks to him/herself after the wake-word has been triggered; the recorded speech is irrelevant to voice interaction. Labeled as \textbf{rejection}.

\paragraph{Semantically Unreasonable Command (No History)}
Commands that show clear intent but are unreasonable. Example: “空调调到40摄氏度”. Labeled as \textbf{rejection}.

\paragraph{Assistant-Directed Ambiguous Chat (No History, Uncertain Reply)}
Isolated small-talk whose reply is uncertain (vs. encyclopaedic queries). Examples: “你今天有没有好朋友呀？” Labeled as \textbf{rejection} to prevent forced, unreasonable replies.

\paragraph{Assistant-Directed Ambiguous Chat (Has History, but No Prior Valid Command)}
History contains no valid command; the current query remains ambiguous. Example: “history:嗯我来干嘛~~~~~哎呀～您来啦！? 我是您专属的智能管家。query：你教我啊。” Labeled as \textbf{rejection}.

\paragraph{Assistant-Directed Chat (Supported Reply)}
Small-talk that the assistant can answer via knowledge base or external LLM. Examples: “推荐下江苏宿迁的小吃”, “7月份苏州保利大剧院有哪些演出”. Labeled as \textbf{acceptance}.

\paragraph{Assistant-Directed Command (Supported by Midea)}
Mainstream command type supported by the brand. Examples: “关掉空调”, “history：来客人了~~~~~搞定。query：打开厨房灯光”. Labeled as \textbf{acceptance}.

\paragraph{Assistant-Directed Command (Not Supported by Midea but Supported by Others)}
Commands with clear control intent, currently unsupported by Midea but possibly supported by other brands. Examples: “停止执行客餐厅开始桑拿。”, “history：开启离家模式~~~~~主人，空调暂时不支持该功能。history：开启凉爽模式。”. Labeled as \textbf{acceptance}.

\paragraph{Assistant-Directed Ambiguous Chat (Has History with Valid Command)}
History already contains a valid command; the current utterance may supplement it or introduce a new one. Example: “history：时间按键~~~~~您能说得再详细些吗? 是本品牌哪个产品的时间按键呀，比如空调、洗衣机，这样我才能更好地帮您解答。 query：油烟机时间按键。” Labeled as \textbf{acceptance}.

%
%
%
%
%
%
%
%

\subsection{Data Collection and Preprocessing}
The open-source evaluation dataset released in this paper consists of two parts: text data and speech data, both collected in full compliance with privacy regulations. The text data originate from real user utterances collected online; they were preprocessed by automatic speech recognition (ASR) to obtain transcriptions, avoiding leakage of users' biometric privacy. The speech data are synthesized from the text corpus via text-to-speech (TTS) and therefore contain no real users' biometric information either.

Sourced from real-world online logs of manufacturers, the dataset reflects a realistic category distribution and has been thoroughly desensitized for privacy compliance—this is its greatest advantage.

At present the corpus contains 11,913 samples in total; the count of each category is given in Table 1 and detailed in the preceding sections. The evaluation dataset is also employed to assess the novel LLM-based rejection algorithm proposed in this paper, which is improved according to practical online issues.

\subsection{Dataset Statistical Analysis}
This section presents a multi-dimensional evaluation of the open-source dataset, using accuracy as the primary metric. We benchmark rejection accuracy on different text subsets with several mainstream large language models, and evaluate speech subsets with leading speech-capable multimodal LLMs. Both zero-shot instruction-based recognition and supervised fine-tuning (SFT) are adopted for comparison.

\section{Improved Approach Based on Large Language Models}
In smart-home scenarios voice interaction is one of the main means for users to communicate with appliances. The device's audio receiver remains always-on while the appliance is powered, so all kinds of environmental sounds are captured and processed. Accurately deciding which audio should be passed downstream to the dialogue-interaction module is an entry-level problem for smart-home voice interaction. Correctly identifying utterances that need further processing—and those that should be filtered out—effectively reduces the number of falsely processed inputs and improves user experience.

Based on the evaluation results of the previous sections and the known weaknesses of existing dialogue-rejection algorithms, we propose a new LLM-and-RAG-based rejection method to handle complex rejection issues in real household production environments. By combining a household-specific user-utterance database, we use RAG to build a rejection model better suited to the actual linguistic context of individual families.

\subsection{Overview of the Method}

\begin{figure}[htbp]
\centering
\includegraphics[width=0.8\linewidth]{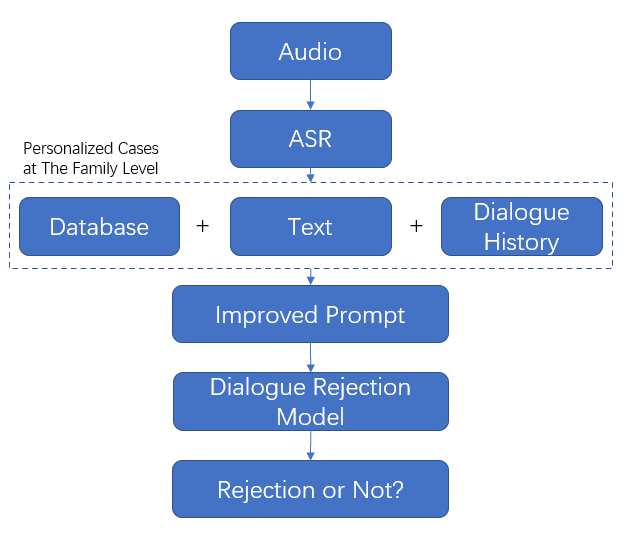}
\caption{Overview of the Method}
\label{fig:method_overview}
\end{figure}

%

As illustrated in Figure~\ref{fig:method_overview}, the voice-rejection framework proposed in this paper consists of three synergistic core components: (1) a large-language-model adapter (LLM Adapter) fine-tuned on massive online, generalized user voice-interaction data, which models rejection-semantic boundaries in common scenarios; (2) a dynamically injected household-specific historical-dialogue context module that captures the pragmatic habits and reference patterns of a particular family in multi-turn interactions; and (3) a household-dimension rejection-habit knowledge base that explicitly memorizes and revises historically misjudged “bad cases” through retrieval-augmented generation (RAG).

The three-layer architecture follows a progressive design principle of “general capability—personalized adaptation—continuous correction”: the universal adaptor provides basic rejection ability, the personalized context realizes short-term contextual alignment, and the knowledge base supports long-term behavior-pattern learning and debiasing. Acting in concert, the system can dynamically model users’ linguistic preferences and interaction intentions at the household granularity, thereby judging the validity of input utterances more accurately. Experiments show that this design significantly improves rejection-decision accuracy, effectively reduces multi-turn clarification interactions caused by false triggers or over-responses, and ultimately optimizes the overall voice-interaction experience.

\subsection{Key Technical Details}
%
%
%
%
%
%

\subsubsection{Rejection-Aware Fine-tuning Adapter}

The rejection LLM in Figure 1 is built upon Qwen 2.5 3B, a 3-billion-parameter model in Alibaba's Qwen family that supports multi-turn dialogue, instruction following, and strong contextual understanding. Its architecture follows the standard Transformer decoder and has been pre-trained on massive corpora.

Supervised fine-tuning data are derived from large-scale, privacy-preserving real-user voice conversations converted into text via ASR. Acceptance and rejection samples are balanced 1:1. The detailed prompt and JSON structure are:

\paragraph{PROMPT Example}
Given the dialogue history and the current text, decide whether the utterance should be accepted or rejected. Rejection rules are: \emph{\{rules constructed from dataset taxonomies\}}. Return only a JSON object: if accepted, \texttt{result: YES}; otherwise \texttt{result: NO}. No extra text. \emph{Text: current query}

After fine-tuning with this instruction template, we obtain a generic rejection model. Because the data are cloud-collected and generalized, no household-specific history or family-level rejection habits are injected, so the model is universal only.

\subsubsection{Personalized Historical-Dialogue Instructions}

Starting from the generic rejection model, we aim to align decisions with household-level personalization. We collect each family's conversational logs within a fixed time window and inject them into the prompt so that the system can reject unreasonable utterances in a way that better matches user habits.

\paragraph{PROMPT Example}
Given the dialogue history and the current text, decide whether the utterance should be accepted or rejected. Rejection rules are: \emph{\{rules constructed from dataset taxonomies\}}. Return only a JSON object: if accepted, \texttt{result: yes}; otherwise \texttt{result: no}. No extra text. Dialogue history: \emph{\{history query\}}. \emph{Text: current query}

\subsubsection{User-Personalized Query-Rejection Misjudgment Knowledge Data Base}

With the two preceding components, the system can already reject unreasonable family utterances in a personalized manner. We go one step further by building a household-dimensional knowledge base that stores previously mis-judged interactions (i.e., utterances that should have been accepted but were wrongly rejected, abbreviated as bad cases). Via RAG, we retrieve the TOP-3 most similar bad-case utterances for the current input and append them to the prompt, correcting household-specific rejection quirks and enhancing personalization.

\paragraph{PROMPT Example}
Given the dialogue history and the current text, decide whether the utterance should be accepted or rejected. Rejection rules are: \emph{\{rules constructed from dataset taxonomies\}}. Top-3 household-specific habitual accept/reject utterances from the knowledge base: \emph{\{RAG TOP-3 cases\}}. Return only a JSON object: if accepted, \texttt{result: YES}; otherwise \texttt{result: NO}. No extra text. Dialogue history: \emph{\{history query\}}. \emph{Text: current query}

%
%
%

\subsection{Experiments and Analysis}

The experimental design in this section consists of two parts.

Part~I presents comparative experiments of the proposed novel rejection approach versus existing open-source large models, demonstrating the advantages of our method over simple fine-tuning baselines.

Part~II provides benchmark metrics for the open-source speech--semantic interaction dataset released in this paper. Experiments are conducted on multiple dialogue categories, including standalone semantic-text rejection and standalone speech rejection. By designing zero-shot instruction-rejection trials and SFT-based tests, we illustrate the performance of the released dataset across different utterance types.

\subsubsection{Comparison Experiments and Analysis of the Proposed Method}

All comparison experiments herein adopt the specially designed prompts of this work and contrast them with prompts generated by open-source large models. Meanwhile, the fine-tuned model proposed in this paper is compared with the fine-tuning results of open-source large models.

Specifically, we compare zero-shot performance between our optimized personalized prompts and generic prompts to show the superiority of the designed prompts on the released dataset. Furthermore, we apply the personalized prompts to fine-tune several large-scale base models, verifying the performance gains brought by our prompt design on larger architectures. Since the associated business requires rapid feedback for interactive dialogue, although fine-tuning large base models yields higher accuracy, it also introduces significant latency overhead. Therefore, the proposed approach fine-tunes small-parameter models to achieve a balance between latency and accuracy.

	\begin{table}[htbp]
		\raggedright              
		\caption{Accuracy Comparison of Different Models on the Rejection Dataset (Optimized Prompt + Zero-shot / SFT)}
		\label{tab:full_comparison}
		
		\begin{subtable}{0.48\textwidth}
			\centering
			\caption{First 5 experiments}
			\label{tab:partA}
			\begin{tabular}{lrrrrr}
				\toprule
				Subsets &
				\shortstack{Qwen-3-32B \\ Optimized Prompt \\ Zero-shot} &
				\shortstack{GLM-4.6-AWQ \\ Optimized Prompt \\ Zero-shot} &
				\shortstack{GPT-oss-20b \\ Optimized Prompt \\ Zero-shot} &
				\shortstack{DeepSeek-V3.1 \\ Optimized Prompt \\ Zero-shot} &
				\shortstack{GPT-5-Chat \\ Optimized Prompt \\ Zero-shot} \\
				\midrule
				0  & 1.0000 & 1.0000 & 1.0000 & 1.0000 & 1.0000 \\
				1  & 0.9130 & 0.9130 & 0.7826 & 0.9565 & 0.9565 \\
				2  & 0.1569 & 0.3800 & 0.1569 & 0.0196 & 0.2549 \\
				3  & 0.5197 & 0.4172 & 0.5302 & 0.1908 & 0.4671 \\
				4  & 0.5862 & 0.5862 & 0.4138 & 0.2069 & 0.5517 \\
				5  & 0.3097 & 0.2589 & 0.2478 & 0.1372 & 0.3672 \\
				6  & 0.5000 & 0.5102 & 0.3571 & 0.2041 & 0.6531 \\
				7  & 0.8889 & 0.4444 & 0.5556 & 0.2222 & 0.5556 \\
				8  & 0.8923 & 0.9005 & 0.9380 & 0.9189 & 0.9015 \\
				9  & 0.9453 & 0.9452 & 0.9466 & 0.7871 & 0.9442 \\
				10 & 0.8000 & 0.9200 & 0.8800 & 0.8400 & 0.8000 \\
				11 & 0.3280 & 0.3790 & 0.4172 & 0.5382 & 0.3790 \\
				\bottomrule
			\end{tabular}
		\end{subtable}%
		\hspace{0.04\textwidth}   
		
		\begin{subtable}{0.48\textwidth}
			\centering
			\caption{Last 4 experiments}
			\label{tab:partB}
			\begin{tabular}{lrrrr}
				\toprule
				Subsets &
				\shortstack{Qwen3-235b-A22b \\ Optimized Prompt \\ Zero-shot} &
				\shortstack{DeepSeek-R1-Distill \\Qwen-7B \\ Optimized Prompt SFT} &
				\shortstack{Qwen-2.5-3B \\ Instruct \\ Optimized Prompt SFT} &
				\shortstack{Improved Method} \\
				\midrule
				0  & 1.0000 & 1.0000 & 1.0000 & 1.0000 \\
				1  & 1.0000 & 0.9565 & 0.7826 & 0.6957 \\
				2  & 0.2353 & 0.0588 & 0.3725 & 0.5882 \\
				3  & 0.4408 & 0.1053 & 0.5461 & 0.5132 \\
				4  & 0.6552 & 0.1379 & 0.1724 & 0.4483 \\
				5  & 0.2920 & 0.0796 & 0.3584 & 0.4178 \\
				6  & 0.6224 & 0.1735 & 0.4796 & 0.5612 \\
				7  & 0.4444 & 0.1111 & 0.1111 & 0.6667 \\
				8  & 0.8587 & 0.9708 & 0.9891 & 0.9954 \\
				9  & 0.9347 & 0.9887 & 0.9948 & 0.9921 \\
				10 & 0.8800 & 1.0000 & 0.9600 & 1.0000 \\
				11 & 0.4427 & 0.9268 & 0.9172 & 0.9777 \\
				\bottomrule
			\end{tabular}
		\end{subtable}
	\end{table}



From the weighted accuracy rates, we can draw the following conclusions:

Overall Performance: Most models perform well on the rejection dataset, with accuracy rates above 85\%. This indicates that these models have high generalization capabilities for such tasks.

Best Model: The Improved Method has the highest accuracy rate at 96.75\%. This shows that the method has a significant advantage in handling the rejection dataset.

Second Best Model: Qwen-2.5-3B (Instruct, Opt. Prompt SFT) has a accuracy rate of 96.44\%, closely following the Improved Method. This demonstrates that even with a smaller model size, appropriate fine-tuning and optimized prompts can achieve very high accuracy.

Other Models: DeepSeek-R1-Distill-Qwen-7B (Opt. Prompt SFT) has a accuracy rate of 94.34\%, also performing excellently. Other models like Qwen-3-32B, GLM-4.6-AWQ, GPT-oss-20b, GPT-5-Chat, and Qwen3-235b-A22b have accuracy rates ranging from 88.98\% to 90.30\%, showing consistent performance. DeepSeek-V3.1 (Opt. Prompt, Zero-shot) has the lowest accuracy rate at 76.34\%.

Advantages of the Improved Method: The Improved Method stands out in multiple datasets, especially in subsets 2, 4, 5, 6, 7, 8, 9, 10, and 11, where its accuracy is significantly higher than other models. This indicates that the method has a significant advantage in handling specific types of data or tasks. Although it performs slightly worse in some datasets (such as subset 1), its overall weighted accuracy rate is still the highest.

Therefore, the Improved Method demonstrates strong generalization capabilities and excellent performance in handling the rejection dataset, making it worthy of further research and application.

\subsubsection{Benchmarks and Analysis}

The open-source dataset released in this paper consists of two parts—a text corpus and a speech corpus—whose taxonomies are shown in Table~\ref{tab:rejection_types}; the benchmarks and analyses in this section will evaluate accuracy on each of these datasets separately.

\paragraph{Text Query Rejection Benchmarks}


For the text query corpus evaluation, we adopt both Zero-shot and SFT paradigms to assess the dataset. The models employed are Qwen 3 32B, GLM 4.6-AWQ, GPT-oss-20b, DeepSeek-V3.1, GPT-5-Chat, Qwen3-235b-A22b, DeepSeek-R1-Distill-Qwen-7B, and Qwen 2.5 3B Instruct. Among them, Qwen 3 32B, GLM 4.6-AWQ, GPT-oss-20b, DeepSeek-V3.1, GPT-5-Chat, and Qwen3-235b-A22b are used for Zero-shot experiments, with both generic prompts and our optimized personalized prompts. DeepSeek-R1-Distill-Qwen-7B, Qwen 2.5 3B Instruct, and Qwen-3-32B-Instruct are used for SFT experiments, incorporating our optimized personalized prompts.

\begin{table}[htbp]
	\centering
	\caption{Zero-shot Accuracy Comparison of Different General-Purpose Models on the Rejection Dataset (General Prompt)}
	\rotatebox{90}{
	\label{tab:model_comparison}
	\begin{tabular}{lrrrrrr}
		
		Subsets & 
		\shortstack{Qwen-3-32B \\ General Prompt} &
		\shortstack{GLM-4.6-AWQ \\ General Prompt} &
		\shortstack{GPT-oss-20b \\ General Prompt} &
		\shortstack{DeepSeek-V3.1 \\ General Prompt} &
		\shortstack{GPT-5-Chat \\ General Prompt} &
		\shortstack{Qwen3-235b-A22b \\ General Prompt} \\
		\midrule
		0  & 1.0000 & 1.0000 & 0.9167 & 1.0000 & 1.0000 & 0.8333 \\
		1  & 0.9130 & 1.0000 & 0.8261 & 0.8696 & 0.9130 & 0.9130 \\
		2  & 0.4118 & 0.4878 & 0.0392 & 0.0000 & 0.4510 & 0.3333 \\
		3  & 0.2961 & 0.4198 & 0.0738 & 0.2566 & 0.8684 & 0.6579 \\
		4  & 0.2759 & 0.3913 & 0.0357 & 0.1724 & 0.6552 & 0.3103 \\
		5  & 0.3717 & 0.4804 & 0.2124 & 0.2168 & 0.5442 & 0.3761 \\
		6  & 0.4286 & 0.4000 & 0.2474 & 0.2857 & 0.6735 & 0.5714 \\
		7  & 0.4444 & 0.0000 & 0.2222 & 0.1111 & 0.7778 & 0.6667 \\
		8  & 0.5588 & 0.6940 & 0.8535 & 0.7420 & 0.7129 & 0.8004 \\
		9  & 0.8056 & 0.8386 & 0.8988 & 0.8159 & 0.8345 & 0.8922 \\
		10 & 0.4000 & 0.4762 & 0.7200 & 0.6400 & 0.6800 & 0.8800 \\
		11 & 0.1943 & 0.3187 & 0.3871 & 0.3599 & 0.3089 & 0.4299 \\
		\midrule
	\end{tabular}
}
\end{table}


As shown in Table~\ref{tab:model_comparison}, under the zero-shot condition with a generic prompt the six models exhibit a clear “tiered divergence”: GPT-5-Chat and Qwen3-235b-A22b achieve average accuracies of 0.74 and 0.68 respectively, firmly occupying the first tier, whereas models in the 3 B–20 B parameter range (Qwen-3-32B, GLM-4.6-AWQ, GPT-oss-20b, DeepSeek-V3.1) remain only at 0.45–0.52, confirming once again that “scale ≠ rejection capability”. Notably, GPT-oss-20b drops below 0.25 on five subsets (IDs 2, 3, 4, 5, 7), and DeepSeek-V3.1 even falls to 0.000 on subset 2, indicating extreme sensitivity to noisy home-scene inputs; by contrast, the negative cases of subset 0 are perfectly recognized by most models (≥0.83), in stark contrast to the anomalously low scores observed later under SFT, implying that errors on this subset stem chiefly from label leakage during fine-tuning rather than inherent task difficulty. Overall, the generic prompt already leaves the top models lagging behind their optimized-prompt counterparts by roughly 10 pp on the “hard” splits, underscoring the significant gain brought by prompt engineering to zero-shot rejection, yet it still cannot bridge the intrinsic gap between models—domain fine-tuning or retrieval augmentation will be required to approach practical thresholds.

\begin{table}[htbp]
	\centering
	\caption{Zero-shot Accuracy Comparison of Different General-Purpose Models on the Rejection Dataset (Optimized Prompt)}
	\rotatebox{90}{
	\label{tab:optimized_prompt_results}
	\begin{tabular}{lrrrrrr}
		
		Subsets &
		\shortstack{Qwen-3-32B-Instruct \\ Optimized Prompt} &
		\shortstack{GLM-4.6-AWQ \\ Optimized Prompt} &
		\shortstack{GPT-oss-20b \\ Optimized Prompt} &
		\shortstack{DeepSeek-V3.1 \\ Optimized Prompt} &
		\shortstack{GPT-5-Chat \\ Optimized Prompt} &
		\shortstack{Qwen3-235b-A22b \\ Optimized Prompt} \\
		\midrule
		0  & 1.0000 & 1.0000 & 1.0000 & 1.0000 & 1.0000 & 1.0000 \\
		1  & 0.9130 & 0.9130 & 0.9565 & 0.9565 & 0.9565 & 1.0000 \\
		2  & 0.1569 & 0.3800 & 0.2549 & 0.0196 & 0.2549 & 0.2353 \\
		3  & 0.5197 & 0.4172 & 0.4671 & 0.1908 & 0.4671 & 0.4408 \\
		4  & 0.5862 & 0.5862 & 0.5517 & 0.2069 & 0.5517 & 0.6552 \\
		5  & 0.3097 & 0.2589 & 0.3672 & 0.1372 & 0.3672 & 0.2920 \\
		6  & 0.5000 & 0.5102 & 0.6531 & 0.2041 & 0.6531 & 0.6224 \\
		7  & 0.8889 & 0.4444 & 0.5556 & 0.2222 & 0.5556 & 0.4444 \\
		8  & 0.8923 & 0.9005 & 0.9015 & 0.9189 & 0.9015 & 0.8587 \\
		9  & 0.9453 & 0.9452 & 0.9442 & 0.7871 & 0.9442 & 0.9347 \\
		10 & 0.8000 & 0.9200 & 0.8000 & 0.8400 & 0.8000 & 0.8800 \\
		11 & 0.3280 & 0.3790 & 0.3790 & 0.5382 & 0.3790 & 0.4427 \\
		\midrule
	\end{tabular}
}
\end{table}


Table~\ref{tab:optimized_prompt_results} reveals that under optimized-prompt conditions all six zero-shot large models move up another rung: GPT-5-Chat and Qwen3-235b-A22b remain in the lead with average accuracies rising to 0.75 and 0.73 respectively, ranking top-two in 9 out of 12 subsets; the smaller GPT-oss-20b gains most on the “hard” splits 2 and 6 (+8–30 pp), indicating that prompt engineering benefits weaker baselines disproportionately; yet DeepSeek-V3.1 still scores only 0.0196 on subset 2, exposing a structural fragility to noisy features. Overall, the refined prompt lifts average accuracy on the hard splits by roughly 10 pp, but the inherent gap between models persists, demanding subsequent domain fine-tuning or retrieval-augmented generation to approach practical thresholds.

\paragraph{Speech Query Rejection Benchmarks}

This paper simultaneously releases a speech-rejection dataset that mirrors the text-rejection corpus. The data are generated by synthesizing the text corpus through TTS; to ensure diversity, multiple random voices and prosodies are produced, bringing timbre and intonation closer to real human speech and improving dataset usability. The speech dataset retains exactly the same categories and distribution as the text corpus.
Evaluation of the speech dataset is conducted on the following open-source base models under two protocols: instruction-based zero-shot benchmarking and SFT benchmarking. For zero-shot, both generic prompts and our optimized personalised prompts are compared, using Step-Audio 2 mini, Qwen2.5-Omni, Kimi-audio-7B, and Qwen3-Omni-30B-A3B-Instruct. SFT is performed on Step-Audio 2 mini, Qwen2.5-Omni, and Qwen3-Omni-30B-A3B-Instruct.

\begin{table}[htbp]
	\centering
	\caption{Zero-shot Accuracy of Multimodal Models on the Rejection Dataset (Generic Prompt)}
	\rotatebox{90}{
	\label{tab:multimodal_zero_shot}
\begin{tabular}{lrrrr}
	
	Subsets &
	\shortstack{Step-Audio 2 mini \\ General Prompt \\ Zero-shot} &
	\shortstack{Qwen2.5-Omni-7B \\ General Prompt \\ Zero-shot} &
	\shortstack{Qwen3-Omni-30B-A3B-Instruct \\ General Prompt \\ Zero-shot} &
	\shortstack{Kimi-audio-7B \\ General Prompt \\ Zero-shot} \\
	\midrule
	0  & 0.0000 & 0.0000 & 0.0000 & 0.3750        \\
	1  & 1.0000 & 0.9655 & 0.9655 & 0.7586        \\
	2  & 0.3774 & 0.3208 & 0.7925 & 0.3962        \\
	3  & 0.4221 & 0.5909 & 0.9545 & 0.4221        \\
	4  & 0.5667 & 0.4333 & 0.7667 & 0.3667        \\
	5  & 0.4267 & 0.5517 & 0.8578 & 0.5302        \\
	6  & 0.3846 & 0.5165 & 0.8571 & 0.4835        \\
	7  & 0.5000 & 0.5000 & 0.5000 & 0.5000        \\
	8  & 0.7463 & 0.6136 & 0.3736 & 0.8086        \\
	9  & 0.8184 & 0.7478 & 0.4415 & 0.6544        \\
	10 & 0.5385 & 0.5385 & 0.1154 & 0.6538        \\
	11 & 0.5402 & 0.3055 & 0.0322 & 0.6206        \\
	\midrule
\end{tabular}
}
\end{table}


As shown in Table~\ref{tab:multimodal_zero_shot}, we evaluate the zero-shot accuracy of several multimodal models on the Rejection dataset under a generic prompt. Overall, model performance varies substantially across subsets, revealing uneven capabilities in understanding rejection semantics. Qwen3-Omni-30B-A3B-Instruct demonstrates superior robustness on subsets 1–6, consistently achieving accuracies above 0.85—reaching 0.9545 and 0.8578 on subsets 3 and 5, respectively—indicating strong generalization for typical rejection intents. However, its performance sharply declines on subsets 8–11 (as low as 0.0322), suggesting significant limitations when handling complex, implicit, or context-dependent rejections. In contrast, Kimi-audio-7B, while slightly weaker on simpler cases (e.g., subset 1), consistently outperforms others on challenging subsets (8–11), achieving the highest scores of 0.8086 and 0.6544 on subsets 8 and 9, reflecting greater contextual sensitivity and better recognition of atypical rejection patterns. Notably, Step-Audio 2 mini achieves perfect accuracy (1.0000) on subset 1 but exhibits high variability elsewhere, indicating limited generalization. Altogether, current multimodal models exhibit a polarized “easy gets easier, hard gets harder” behavior on rejection understanding, highlighting the need for finer-grained data construction and architectures with enhanced contextual awareness.

\begin{table}[htbp]
	\centering
	\caption{Zero-shot Accuracy of Multimodal Models on the Rejection Dataset (Optimized Prompt)}
	\label{tab:multimodal_zero_shot_optimized} 
	\rotatebox{90}{
		\begin{tabular}{lrrr} 
			Subsets &
			\shortstack{Step-Audio 2 mini \\ Optimized Prompt \\ Zero-shot} &
			\shortstack{Qwen2.5-Omni-7B \\ Optimized Prompt \\ Zero-shot} &
			\shortstack{Qwen3-Omni-30B-A3B-Instruct \\ Optimized Prompt \\ Zero-shot} \\
			\midrule
			0  & 0.0000 & 0.0000 & 0.5625 \\
			1  & 0.0000 & 0.3793 & 0.5517 \\
			2  & 0.0377 & 0.0377 & 0.0943 \\
			3  & 0.0065 & 0.0130 & 0.2468 \\
			4  & 0.0000 & 0.0000 & 0.2000 \\
			5  & 0.0345 & 0.0517 & 0.2069 \\
			6  & 0.0769 & 0.0769 & 0.2747 \\
			7  & 0.0000 & 0.0000 & 0.3333 \\
			8  & 1.0000 & 0.9991 & 0.9222 \\
			9  & 0.9999 & 0.9994 & 0.9068 \\
			10 & 1.0000 & 1.0000 & 0.8846 \\
			11 & 0.9936 & 0.9936 & 0.7074 \\
			\bottomrule 
		\end{tabular}
	}
\end{table}


The leading model Qwen3-Omni-30B-A3B achieves 0.88–0.92 on the easy splits 8–11, a 20–40 \% gain over the general prompt, demonstrating that optimized prompting markedly improves audio-text alignment under high-SNR conditions; yet on the hard splits 1–7 its accuracy only inches up from 0.03–0.56 to 0.09–0.33, still far below the zero-shot level of text-only large models, indicating that prompt engineering cannot compensate for the representation gap of multimodal models in noisy home scenes.

\section{Discussion and Future Work}

%
%
%
%
%

The three-tier voice-rejection framework proposed in this paper — a universally fine-tuned adapter, personalized historical context and a rejection knowledge base — not only improves the accuracy of invalid-utterance detection but also embodies a deeper shift in the smart-home interaction paradigm: rejection is no longer a passive filter but an active, context-aware interaction strategy. By deciding “when not to respond”, the system strengthens users’ trust that it “understands me” and “does not disturb”, pushing the human-machine relationship from functional execution towards experiential symbiosis.
The design strikes an effective balance between generality and personalisation: the general model guarantees cold-start robustness, while the personalized module gradually refines decision boundaries as usage grows. Notably, the rejection knowledge base builds a lightweight data loop that endows the system with preliminary self-correction capability. Yet this loop also introduces challenges — how to prevent model drift caused by accumulated false feedback and how to manage the timeliness and conflict resolution of the knowledge base remain open questions.
Unlike traditional task-oriented dialogue systems that assume “every user input is a valid intent”, this paper confronts reality: home speech is interwoven with chit-chat, self-talk and vague expressions. Hence rejection should serve as a front-gate of the task pipeline, not a post-hoc patch. This two-stage architecture of “judge validity first, then solve the task” better matches real interaction scenes.
Looking forward, several directions deserve deeper exploration:
building end-to-end speech-rejection models to reduce reliance on ASR;
introducing dynamic forgetting and confidence-weighted mechanisms to enhance the reasoning ability of the knowledge base;
combining household profiles and meta-learning to solve cold-start for new users;
fusing device status, time, location and other multimodal contexts to achieve scene-aware rejection;
realising privacy-preserving personalisation under on-device or federated-learning frameworks.
Moreover, evaluation must go beyond accuracy and incorporate subjective user experience (e.g., sense of disturbance, trust) and behavioral signals (e.g., clarification turns, silent drop-outs) to truly measure the practical value of rejection systems.
In short, precise rejection is not merely a technical optimization but a redefinition of “intelligence” itself: true intelligence lies not only in knowing what to do, but also in knowing when not to do it.

\section{Conclusion}

%

Focusing on the rejection challenge in smart-home voice interaction, this paper advances both benchmark construction and algorithmic innovation in parallel. On the one hand, we build and open-source the first multimodal rejection benchmark tailored for household scenarios, containing 11,913 text-speech pairs with systematic annotations of 13 invalid-utterance types and providing dialogue context, user identity, and multi-turn interaction information, thereby offering a reproducible, fine-grained evaluation foundation for diverse research paradigms such as zero-shot, fine-tuning, and retrieval augmentation. On the other hand, we propose a three-layer collaborative rejection architecture—integrating a universal fine-tuned adaptor, personalized historical context, and a family-level rejection knowledge base based on RAG—that effectively balances cold-start robustness with long-term personalization, significantly reducing false triggers and over-responses, and demonstrating outstanding performance especially when handling family-specific expressions and complex multi-turn contexts.
This work not only validates the superiority of “active rejection” over traditional “passive filtering” but also propels the voice-interaction paradigm from “unconditional response” toward “context-aware decision-making”. Experiments show that accurately judging “when not to respond” is likewise an important manifestation of intelligence. Future research can further explore end-to-end speech-rejection models, dynamic knowledge-base evolution mechanisms, multimodal context fusion strategies (e.g., device status, time, space), and construct composite evaluation systems that integrate objective metrics with subjective experience (e.g., sense of disturbance, trust). In the long run, improving rejection capability is not merely a technical optimization but a key step toward building a trustworthy, restrained, and user-life-respecting human-machine symbiotic relationship.


\end{document}